# Search for $\theta_{13}$ mixing parameter with the low power reactor

V.N. Kornoukhov[1], A.S. Starostin[2]

(RSC ITEP, Moscow)

## Abstract

One of the main problems of neutrino physics is the measurement of mixing parameter $\sin^2 2\theta_{13}$. At present time several projects of reactor experiments with sensitivity of $\sin^2 2\theta_{13} \approx 0.01$ are under discussion. Practically, all of them are based on the principle "one reactor- two detectors". Such approach supposes to use NPP reactor with power of several GW as the source of neutrino and two detectors of similar configuration placed at different distance from the reactor. The systematic error in such experiments may reach ~ 1% and the accuracy ~ 0.01 as the result.

In this paper we propose to use for the measurement of $\sin^2 2\theta_{13}$ the existing SuperKamiokande (SK) set-up in combination with "own" antineutrino source – nuclear reactor of small thermal power ~ 300 MW (LPR). Such experiment could be carried out in relatively short time. As it was shown by the study of different detection mechanisms, the combination LPR-SK allows to obtain the sensitivity of $\sin^2 2\theta_{13} \approx 0.002$.

---

[1] E-address: kornoukhov@itep.ru
[2] Corresponding author: starostion@itep.ru

Introduction

The study of the mass structure of neutrino is considered one of the general goals in neutrino physics. The effective tool to resolve this problem is the experiments on neutrino oscillations search. The observation of neutrino oscillations gives the opportunity to determine the parameters of the Pontecorvo - Maki - Nakagawa - Sakata (PMNS) matrix [1,2] – the mixing angles $\theta_{12}$, $\theta_{23}$, $\theta_{13}$, and the mass square differences $\Delta m^2_{21} = \Delta m^2_2 - \Delta m^2_1$, $\Delta m^2_{32} = \Delta m^2_3 - \Delta m^2_2$, $\Delta m^2_{31} = \Delta m^2_{21} + \Delta m^2_{32}$. The full information on matrix parameters allows to determine the structure of active neutrino, in particular, electron neutrino $\nu_e$ could be presented as a superposition of neutrino mass states ($\nu_1$, $\nu_2$, $\nu_3$) with certain weight factors:

$$\nu_e = \cos\theta_{12}\cos\theta_{13}\, \nu_1 + \sin\theta_{12}\cos\theta_{13}\, \nu_2 + \sin\theta_{13}\, \nu_3. \quad (1)$$

During last years a set of positive results in experiments on the search of atmospheric, solar and reactor neutrino oscillations has been obtained. The joint analysis of experimental data from SK [3] and K2K [4] experiments (atmospheric neutrino), CHOOZ [5], Palo Verde [6] and KamLAND [7] experiments (reactor neutrino) and all data on solar neutrino [8], including the results from SNO [9], has allowed to draw conclusions concerning the mechanism of oscillations and, partly, determined the parameters of PMNS matrix. In the assumption of natural neutrino mass hierarchy ($m_1 < m_2 < m_3$) the results of joint analysis of these experiments are represented as follows [10,11]:

$$\Delta m^2_{sol} \equiv \Delta m^2_{12} = 7.1\,^{+1.2}_{-0.6} \cdot 10^{-5} \text{ eV}, \qquad \sin^2 2\theta_{12} = 0.821\,^{+0.064}_{-0.062}\,;$$

$$\Delta m^2_{atm} \equiv \Delta m^2_{23} \approx \Delta m^2_{13} = 3\,^{+3}_{-2} \cdot 10^{-3} \text{ eV}, \qquad \sin^2 2\theta_{23} = 1\,_{-0.2}\,. \quad (2)$$

In presented data there is lack of very significant element of the PMNS matrix, namely, the mixing angle $\theta_{13}$. At present only the limit on this parameter has been obtained from the results of the reactor experiment CHOOZ [5]:

$$\sin^2 2\theta_{13} \leq 0.14 \text{ (90\% C.L. at } \Delta m^2 = 2.5 \cdot 10^{-3} \text{ eV}^2). \quad (3)$$

However this is not enough to solve a set of important problems of fundamental physics. Besides the problem of reconstruction of the structure of active neutrino (1), the presence of nonzero $\theta_{13}$ parameter is the necessary condition for the existence of CP - violation in lepton sector. Therefore, the establishment of nonzero value of $\sin^2 2\theta_{13}$ or improvement of its limit at least by the order is extremely important. Furthermore, there is opinion that if $\sin^2 2\theta_{13} < 0.01$, electron-neutrino must have a different flavor quantum number from muon and tau neutrinos and a new quantum number requires a new symmetry [12].

To resolve this problem the large scale (both on amount of efforts, and at cost) accelerator experiments with standard neutrino beams (К2К [4] and MINOS [13] experiments), and also with so-called "super beams" (JHF/Super-Kamiokande [14] and NuMI [15] experiments) were proposed. However, it was demonstrated by careful analysis [16,17,18] that the preferable way to achieve the maximum sensitivity of the $\sin^2 2\theta_{13}$ parameter is the reactor experiments. High sensitivity of the reactor experiments could be reached at relatively low cost and term of measurements compared with the accelerator ones.

1. Reactor experiments.

For the search of oscillations of reactor antineutrino the inverse beta-decay reaction is used:

$$\bar{\nu}_e + p \rightarrow e^+ + n \qquad (4)$$

In this process, practically, all neutrino energy, excluding reaction threshold – 1.804 MeV, is transmitted to positron. After the capture of annihilation gammas the energy equal to (2 x 0.511) MeV is added to the total transmitted energy. Therefore the minimum energy of the spectrum from reaction (4) in the scintillator detector, for which own threshold does not exceed 100 - 200 keV is equal to ~1 MeV. The threshold of the Cherenkov detector, which is defined only by positron kinetic energy is much higher.

The problem of separating the spectrum (4) from background can be well simplified due to the method of delayed coincidence between the positron and the neutron signals. The time interval between these events depends on the properties of the detecting substance (namely, the thermalization and capture constants). The time delay between the positron event and neutron capture in water is about 1 ms, with the energy release of 2.2 MeV. In the case of gadolinium dissolved into water, the time interval is reduced up to ~ 100 μs and the total energy of γ-cascade following the neutron capture is about 8 MeV.

The reactor oscillation experiments are based on the "disappearance" principle when electron neutrino changes its flavor ($\nu_e \rightarrow \nu_{\mu,\tau}$) and could not manifest itself in the inverse beta - decay reaction. This results in the change of the count rate of the detector and the distortion of the positron spectrum (4). In the case of mixing of two mass states $\nu_1$ и $\nu_3$ the probability of transition $\nu_e \rightarrow \nu_\tau$ is defined by the expression:

$$\sim \sin^2 2\theta_{13} \cdot \sin^2(\varphi), \qquad (5)$$

where $\sin^2 2\theta_{13}$ stands for the mixing parameter, and $\varphi = (1.27 \cdot \Delta m^2_{31} \cdot L \cdot E^{-1})$. $\Delta m^2_{31}$ stands for the mass squared difference of $m^2_3$ and $m^2_1$, E is the energy of reactor antineutrinos, and L is the distance from the reactor to the detector. A value of $\Delta m^2_{31} = 3^{+3}_{-2} \cdot 10^{-3}$ eV is known from the analysis of the results [10, 11]. The optimum distance L for the observation of ($\nu_e \rightarrow \nu_\tau$) transition is defined from the condition of reaching the first oscillation maximum in the expression (5) ($\varphi \approx \pi/2$). Taking into account the value of known parameters, L is equal to ~ (1.5 – 2.0) km.

It is possible, that the value of $\sin^2 2\theta_{13}$ can be much less than the existing experimental limit [5,19]. Therefore, the required accuracy of new generation of the reactor experiments on the mixing angle $\theta_{13}$ measurements should be of an order of 1%. The suitable statistical accuracy can be obtained due to increase of the detector mass. The target of the CHOOZ detector had the mass of 5 tons, and the statistical error was equal to 2.8%. Therefore the accuracy $\leq$ 1% can be reached when the mass of the detector will be over 50 t. More difficult problem is to reduce the systematic error of measurements. It comes from various sources: i) an uncertainty of the reaction cross-section (4); ii) reactor burn up effects and respective change of a neutrino flux; iii) uncertainties of detector efficiency and number of protons in target. The systematic error of the CHOOZ experiment was 2.7%. For its reduction the authors of the Kr2Det project [16] have proposed the so-called "one reactor - two detectors" method, which is based on the use of two *identical* detectors with the same mass and design. One of them is a "near" detector, placed at the distance of 100 – 150 m from the reactor to avoid oscillation effect. It serves for the measurement of undistorted positron spectrum. The second one is a "far" detector, placed in the area of the most effective oscillation at the distance of 1500 – 2000 m from the reactor. The oscillations will be detected and their parameters could be obtained from the comparison of the positron spectra from "near" and "far" detectors. At that the results of the analysis [16] do not depend on the exact knowledge of the reactor parameters, the antineutrino spectrum, concentration of atoms of hydrogen, detector efficiency and effective volume. Thus the main sources of a systematic error are eliminated and, as calculations have showed, the systematic error will not exceed 0.5% for such type of experiments [16,17].

The detailed analysis of various approaches for reaching the sensitivity of $\sin^2 2\theta_{13} < 0.01$ in reactor experiments has been done in [18]. The authors also hold to the scheme of experiment with "near" and "far" detectors. As a basic characteristic of a reactor experiment it was considered the integrated luminosity *L*, which was expressed in units of detector mass [tons] x thermal power of reactor [GW] x running time [years]. It has allowed to analyze the sensitivity for two categories of the experimental benchmark setups: Reactor - I, with a luminosity $L_1$ = 400 t·GW·y and Reactor-II, with a luminosity twenty times more, $L_2$ = 8000 t·GW·y.

Two types of systematic errors were introduced: $\sigma_{norm}$ - the total error, including the uncertainties of parameters of the antineutrino flux and the parameters of near and far detectors; $\sigma_{cal}$ – the error of detector energy calibration. Besides that a number of conditions concerning the characteristics of the detectors was assumed:

- ❖ to reduce the systematic error the near and far detectors should be completely identical (except their volumes only);
- ❖ the mass of "near" detector should be ten times less than the mass of "far" detector, and it is placed at the distance ~ 100 - 150 m from the reactor, in "oscillations-free" zone. The distance between the reactor and the "far" detector is ten times more. As a result, the counting rate of "near" detector should exceed by the order of magnitude the counting rate of "far" detector.

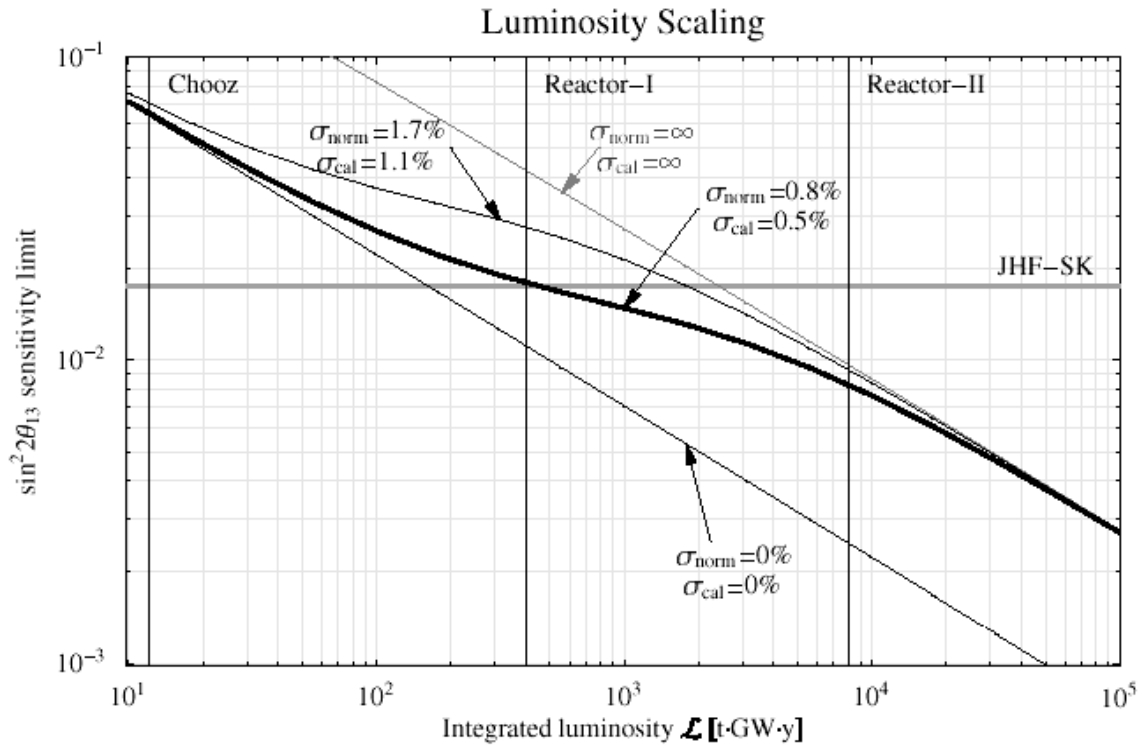

Fig.1. The sensitivity to $\sin^2 2\theta_{13}$ as a function of integrated luminosity for different values of the normalization error $\sigma_{norm}$ and the energy calibration error $\sigma_{cal}$ at the 90% confidence level [18].

As a result of the analysis the functional dependence of $\sin^2 2\theta_{13}$ sensitivity limit on the integrated luminosity was obtained. As it is shown in Fig.1 for ***L*** > $10^4$ t·GW·y the experiment sensitivity is plotted as a ***$L^{-1/2}$*** function and weakly depends on systematic errors.

The authors explain the decrease of influence of systematic errors on the sensitivity by the fact that due to big statistics (L > $10^4$ t·GW·y) it is possible to measure simultaneously the $\sin^2 2\theta_{13}$ parameter *and* the normalization with high accuracy (another words, the determination of the normalization by "far" detector itself becomes more accurate than the external input $\sigma_{norm}$).

The general conclusion that follows from [18] is the possibility to reach the sensitivity to $\sin^2 2\theta_{13}$ of about $10^{-3}$ in the experiments with integrated luminosity of ~ $10^5$ t·GW·y. To reach such a scale of ***L*** one should use the "near" and "far" detectors with mass of $10^3$ tons and ~$10^4$ tons respectively, the running time of measurements is about 5 years and the thermal power of a reactor is 3 GW. It means that the "near" detector should have the mass of the KamLAND detector. Nevertheless, from engineering and technological point of view the design of such detectors is technically available.

However, the experimental complex of such a scale practically under the operating nuclear plant is a big problem because of engineering and civil safety. Even if this problem will be resolved it will demand additional cost and time. Therefore, for reaching the luminosity ***L*** = $10^4$ - $10^5$ t·GW·y and correspondent sensitivity of $\sin^2 2\theta_{13}$ < 0.01 we propose another approach, namely, using a detector with mass of ≥ $10^4$ t (the "far" detector) in conjunction with its "own" antineutrino source. A nuclear reactor of relatively small thermal power ~ 300 MW is proposed to be used as such a source. Such experiment can be realized in rather short time if one used the small power reactor (SPR) together with the existing SuperKamiokande detector. As it will be shown below, the combination "SPR – SK" allows to reach the luminosity ~ $10^5$ t·GW·y correspondent to the sensitivity to $\sin^2(2\theta_{13})$ parameter ~ 0.002. Further we will hold to the following plan. In chapter II we shall present the general scheme of LPR – SK experiment, including the experiment geometry and general characteristics of the LPR, "near" and "far" detectors. In chapter III possible methods of reactor antineutrino detection by the SK will be discussed. In chapter IV the estimations of the sensitivity on $\sin^2 2\theta_{13}$ parameter in the SPR – the SK experiment will be presented.

## 2. The scheme of experiment.

We shall follow to the scheme of the "one reactor - two detectors" experiment [16]. A near (ND) and a far (FD) detectors are completely identical ones, except their volumes. The "near" detector is placed at the distance ~ (100 ÷ 150) m from the reactor, in "oscillation-free" zone, at the depth of ~ 250 m.w.e. The SuperKamiokande [3,20] (the "far" detector) is located 1000 m underground under mountain Ikenoyma, at a height of 370 m above sea level (Fig.2). This depth corresponds to effective thickness of 2700 m.w.e. The shortest distance across from the underground laboratory to the surface is 2.3 km. Such a profile of the mountain allows to choose the optimum distance between the reactor located on a slope of the mountain, and the "far" detector. As follows from (5), the first oscillation minimum L depends on the average neutrino energy and, hence, on the threshold of the detector. Below various ways for lowering of the SK threshold will be discussed. Now, for distinctness, we fix the threshold equal to 3.8 MeV for reactor antineutrinos. For this energy the optimum value L is about 2 km. To fit the experiment conditions formulated in [16] the counting rate of the ND detector should exceed an order of magnitude of the count rate of the FD detector. Therefore, taking into account the distances between the reactor and "far" and "near" detectors, the ND should have a mass which is ~ 20 times less than the mass of the SK.

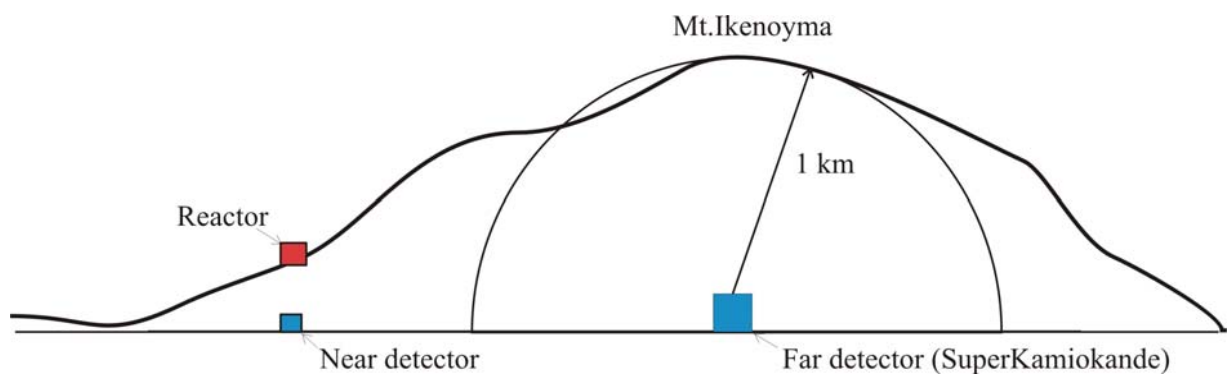

Fig.2 A profile of the Mount Ikenoyma along the direction of smallest extent with location of SuperKamiokande, the near detector and reactor.

## 2.1. The small power reactor.

As an antineutrino source it's proposed the KLT-40S nuclear setup. This system consists of two nuclear reactors with the total thermal power of 300 MW (2 x 150 = 150 MW), equipped with the cooling system, the safety system, and the storage of the spent nuclear fuel [21]. Since 1959, such reactor setups are operated in Soviet Union (nowadays the Russian

Federation) as power plants of the icebreakers. Now the project of floating thermal power station (FTPS) based on the KLT-40S is developed. It's planed to use FTPS in the north and eastern regions of the Russian Federation [22]. There is also a project to cite such FTPS near the coast in China [23].

The water-water type reactors are placed inside a tank with the metal/water shielding and equipped with a safety system, which satisfies all the standards and recommendations of the IAEA (the International Atomic Energy Agency). The equipment of the reactors and the control systems are located in a pressurized casing. The water consumption of the reactor cooling system is $2 \times 250 = 500$ m$^3$/hour. The reactor campaign is about 3 years. The KLT-40S meets the most rigid requirements on seismology (permissible acceleration is equal to 12 g). The minimal volume needed for only to mount the reactors is equal to $25 \times 25$ (h) $\times 30$ m$^3$. The dimension of the reactor case is $\varnothing 2.93 \times 5.35$ m with the mass of 62 tons. The complete set of an atomic power station with two steam turbines and two generators is placed in a volume of $25 \times 25 \times 70$ m$^3$. The atomic power station based on the KLT-40S reactors can generate 520 GW hour of electricity per year. At the price of 0.1 \$/kW·h, the total cost of the electricity generated during one year is about 50 M \$ [24].

It is necessary to note one more advantage of the experiment with the "own" reactor. Reactors of the KLT-40S type are easy handled and low inertial. In the framework of the scientific experiment there is an opportunity to set an optimum ratio in duration of reactor "on" and reactor "off" modes. That allows to perform background measurements with high accuracy.

## 2.2. The far detector - SuperKamiokande.

The SuperKamiokande detector is designed to study solar and atmospheric neutrino, and also such exotic processes, as proton decay. The SK is a water Čerenkov detector and has a cylindrical geometry, 39.3 m in diameter and 41.4 m in height. Within the tank, a stainless steel framework supports separate arrays of inward-facing and outward-facing PMTs. The inward-facing PMTs, and the volume of water they view, are referred to as the inner detector (ID). The mass of the fiducial volume of ID is equal to 22 500 tons. The outer detector serves as a veto detector for incoming cosmic rays and muons as well as a shield layer against γ-rays and neutrons coming from the rock. Neutrino events, which occur in the ID are reconstructed by using the charge and timing data from the hit PMTs (at the absence of a signal in the external detector). The energy resolution for 4.89 MeV electrons is 20.9%. The minimal registered energy is equal to 3.5 MeV, and the analysis threshold was set at 4.5 MeV. These data are related toward the end of the phase SK-I, just before the failure in the autumn of 2001.

One of key parameters of the reactor experiment is the energy threshold of the detector, $E_{th}$. Its value determines both the range of detected antineutrino spectrum and its average energy, $\bar{E}_\nu$. Those, in tern, define the optimum distance L between the reactor and the detectors and experimental statistics. In solar neutrino experiment the SK detected electrons scattered by neutrinos with the analysis threshold $E_{th} \geq 4.5$ MeV. The value of the threshold is determined by PMTs' sensitivity, and by the sharp increase in background at the lowering of energy threshold.

Keeping of such energy threshold (~ 4.5 MeV) for detection of positrons in reactor experiment it will result in the big loss of useful events. Moreover, taking into account the Q-value for the inverse beta-decay reaction (1.8 MeV), the eventual threshold for antineutrino will be 6.3 MeV. More important sequence lies in the fact that we will not be able to detect the neutron capture in water because of low energy of the accompanying γ-rays ($E_\gamma$ = 2.2 MeV). Thus we cannot apply the coincidence technique to distinguish useful events from reaction (4) and from the background.

The high threshold is a "universal" disadvantage of the SK, which restricts its opportunities, practically, in any experiment. This is why, during the reconstruction of the detector after the incident, modification of hardware and software is underway. Toward the end of these works the SK detector will register recoiled electrons with the kinetic energy of 2 MeV [25]. It will correspond to the threshold for reactor antineutrino of 3.8 MeV. The detector efficiency will be at the level of 60% for the positrons from the reaction (4). To reduce the Rn background, new PMTs have been covered with the protective film, which keeps the radon emanation out of the sensitive volume of the detector [26].

## 2.3. The near detector.

According to the general scheme of the experiment, the near detector (ND) should be a full analogue of the far detector, namely, the SK. It will be placed at the distance of ~ 100 m from the reactor, in an "oscillation-free" zone, at the depth of about 250 m.w.e. Its count rate should exceed by an order of magnitude the count rate of the far detector. Taking into account this assumption and, also the distance between detectors and the reactor, the fiducial volume of the ND should be ~ 20 times less than that of the SK. At the depth of 250 m.w.e., the count rate of the active veto system from the charge component of the cosmic ray will be about 300 events/s. At the duration of the anticoincidence signal ~ 100 μs, such count rate of veto system will not effect the registration efficiency of the detector.

3. The possible methods of reactor antineutrinos detection.

Several methods of detection of reactor antineutrinos can be used in the SPR - SK experiment on the mixing angle $\theta_{13}$ search:

1. The SK is operated in its standard mode as a water Cherenkov detector. In this case only positrons will be registered.

2. Gadolinium salt is dissolved in the water of the SK tank. The Cherenkov mode of the detection remains, but together with positron, delayed signal from captured neutron is registered.

3. As the detection medium a scintillator is applied: the SK becomes a scintillator detector. Positrons and neutrons are registered.

Let's consider in details the parameters of the experiment and achievable sensitivity of $\sin^2 2\theta_{13}$ measurements for each of the listed detection modes.

*3.1. A standard mode of the water Cerenkov detector.* To detect the solar neutrinos with the SK detector one used the $\nu_e$ - e elastic scattering reaction with the threshold of 4.5 MeV. Scattered electrons have anisotropic distribution with forward peak into the $60^o$ - cone of the same direction as the neutrinos (average energy of the boron solar neutrino is ~ 8 MeV). This angular asymmetry is used to separate solar events from background.

In case of the reactor experiment on the measurement of $\sin^2 2\theta_{13}$ parameter we assume that the low threshold of the detector is equal to $T_{kin} \geq 2$ MeV for the kinetic energy of positrons [25, 26]. This threshold is expected for the phase SK-III, and it will correspond to the antineutrinos energy $E_\nu \geq 3.8$ MeV. But simultaneously with an increase of useful statistics, this threshold will provoke sharp increase of the detector background. If proceed from the SK-I results on the background measurements at thresholds 4.5, 5.0, 5.5 and 6.0 MeV [27], than one must to expect the background rate increasing by about three orders following the reducing of the threshold from 5 MeV up to 2 MeV. By our estimations, in this case the SK background will be about $6 \cdot 10^4$ events/day, mainly due to the fact that Rn is dissolved in the water of the detector [27]. This background cannot be decreased neither due to the asymmetry of positron angular distribution in reaction (4) nor due to the method of delayed coincidence between signals from positron and neutron. Positrons after the inverse beta-decay reaction (4) will have practically isotropic distribution. To use the delayed coincidence method it is necessary to detect both products of reaction (4) - positron and neutron. The capture of a neutron in water leads to the emission of γ quantum with the energy of 2.2 MeV followed by several Compton electrons. Therefore, taking into account the threshold of the Cerenkov detector ($T_{kin} \geq 2$ MeV), the most part of Compton electrons will not be registered.

The only way to select the positrons over background with *standard water Cerenkov detector* is using the reactor on/off measurements. During the reactor -"on" phase the useful and background events are collected. Then, during the reactor -"off" period the measurement of a background is carried out. The effect is from subtraction ($N_{eff} = N_{"on"} - N_{"off"}$), and an accuracy of measurements is defined by a statistical error of the difference [$\Delta N_{eff} = (N^2_{"on"} + N^2_{"off"})^{0.5}/N_{eff}$]. The total time of experiment may come to 5 years, for 2.5 year of the reactor off/on measurements. At the absence of oscillations the number of registered positrons ($N_e$) is defined by the following parameters: thermal power of the reactor (0.3 GW), a distance between the reactor and the far detector (2.0 km), the cross-section of the inverse beta-decay reaction, a fiducial volume of the detector (~20 Kt) and its threshold (3.8 MeV). In view of listed key parameters of the experiment, the detector count rate $N_e \sim 800$ events/day, that is almost by two order less than the expected background. By our estimations, at such a ratio of effect/background the statistical error of the experiment will be ~1.4%, that is essentially worse than the statistical error in [16,17,18]. There are two opportunities for increasing the statistical accuracy up to value of about 0.1%:

1. To decrease the background by two orders of magnitude. As it has been mentioned above, works in this direction have been already carried out;

2. To lower the threshold for positrons up to $T_{kin} \geq 1.5$ MeV. In this case one can detect a neutron (by the method of delayed coincidence) and, as a consequence, background will be decreased approximately by three orders of magnitude.

*3.2. The Cerenkov detector with a water solution of gadolinium salts.* Now, the opportunity to use the SK detector for the registration of neutrino from far reactors is being discussed [26,27]. To detect the neutrons, the water solution of Gd salts is proposed. As it was already mentioned, the delay time between signals from positron and neutron in this case is ~ 100 μs, and the total energy of γ-cascade after neutron capture is ~ 8 MeV. According to the calculations [28], the visible energy for these gammas in the Cerenkov detector is (5 ± 2) MeV. To absorb practically all neutrons after the inverse beta-decay reaction (> 90%), a Gd-concentration in water should be at the level of ~ 0.1 % of mass (it means about 100 tons of Gadolinium chloride, $GdCl_3$). The criterion how to recognize reactor antineutrinos is two events occurred in the same area (R < 3 m) within the time interval of ~ 100 ms. The first one is a positron with energy $T_{kin} \geq 2$ MeV, the second one is from the gamma-cascade with average visible energy of (5 ± 2) MeV. Simple estimations show, that it is enough to suppress all kinds of background by 8 orders of magnitude. The only source of background, which can influence the accuracy of the experiment, is the flux of antineutrinos from the far reactors. The average distance from Kamioka laboratory up to the surrounding nuclear stations is ~ 175 km, and their total thermal power is ~ 70 GW [7]. The antineutrino count rate of the SK from far reactors is about 17 events/day, or ~ 2% of total detector's rate. According to the results from the reactor experiments (Rovno [29], CHOOZ [5], KanLAND [7]), the thermal power of nuclear stations is controlled within 2 % of the accuracy. Thus, a neutrino flux from the far NPP's is known at the same level of accuracy. Therefore, the contribution to the error of measurement from neutrino from far NPP's will be negligible. Nevertheless, measurements of the antineutrino from the distant NPP's have its own value. In SPR - SK experiment they can be used for the specification of KamLAND results on the measurement of $\Delta m^2_{12}$ mass parameter.

Thus, the statistical accuracy in the SPR – SK experiment with the SK detector doped with Gd salt weakly depends on background and may reach ~ 0.1%.

*3.3. Scintillation mode of the detection.* The most radical solution to increase the sensitivity of measurements is to use the SK as a scintillation detector (SD). It will lead up to a number of positive changes in comparison with above mention modes. The energy threshold of SD will allow to measure a spectrum of positrons of the inverse beta decay reaction (4) starting from ~ 0.8 MeV and, accordingly, a spectrum of the reactor antineutrino from the energy of 1.8 MeV. The first consequence is the increase of antineutrino flux by 1.7 times. The second one is the correspondent reduction of average antineutrino energy and the change of optimum distance "the reactor - the far detector" from 2 km to ~ 1.5 km. In turn, it will lead to new increase of the antineutrino flux in 1.4 times.

The background will depend on the purity of scintillator. Taking into account the expected count rate of useful events in the far detector (~ 1900 events/day), the requirements to a level of impurity of long-living radioactive elements in scintillator can be not so strong, in comparison with the KamLAND experiment. The extra way to decrease the background is to use a scintillator doped with gadolinium. A time delay between signals from positron and neutron will be reduced in ten times, that will lead to corresponding reduction of background of accidental coincidences. It seems that while using gadolinium the level of background will be low enough not to influence the statistical error of measurements.

## 4. Achievable sensitivity on $\sin^2 2\theta_{13}$ parameter.

When we investigated various methods of reactor antineutrino detection, we remained in the frames of main demands to the "one reactor - two detectors" experiments. It allows to use the results of analysis [18] and to define the achievable limit in measurements of $\sin^2 2\theta_{13}$ via the integrated luminosity of experiment $\mathcal{L}$, where $\mathcal{L}$ = mass of the detector [tons] x thermal capacity of reactor [GW] x duration of measurements [year]. Nevertheless, there are two differences between the parameters of the experiment mentioned in present paper and the parameters specified in work [18], which must be taken into account. These are the distance between the reactor and the far detector as well as the energy range of reactor antineutrino. According to the definition of luminosity in [18] the distance and the energy range were accepted to be 1.7 km and 1.8 MeV $\leq E_\nu \leq$ 8 MeV respectively.

The key parameters of reactor experiment on $\sin^2 2\theta_{13}$ measurements using the low energy reactor are listed in Table I for various modes of antineutrino detection. The luminosity, the number of registered positrons and the sensitivity of experiment were calculated using these data. For Water Cerenkov Detector (WCD) we will consider only cases I and II which provide statistical accuracy at the level of ~ 0.1%. Case I corresponds to the on/off -mode with the level of background two orders lower than the present background of SK. For case II the energy threshold for positrons is about $T_{kin} \geq 1.5$ MeV, that allows to use the method of delayed coincidence and as a consequence, to decrease the background by three orders of magnitude. The original value for the calculation of luminosity $\mathcal{L}$ was taken as 22 500 tons x 0.3 GW x 5 years (33 750 t·GW·y). Then it has been corrected by the value of the reactor - detector distance, the time of the "reactor on" measurement, the detector threshold for antineutrino spectrum, and neutron detection efficiency of the detector. After that the luminosities obtained in this paper and [18] correspond to the same number of registered positrons and, as a consequence, to the same statistical accuracy. The distance from the set-up to the reactor can be obtained from the formula (5). The antineutrino energy is calculated as the average value dependent on the energy threshold of the set-up. Such approach can be justified for the experiments with relatively low statistics ($\mathcal{L} \approx 10^3$ t·GW·y), when it is necessary to use wide range of antineutrino spectrum for the selection of oscillation effect.

Table I

| Detector mode | Distance [m] | Period"on" [days/year] | Threshold [MeV] | $\mathcal{L}$ [Kt×GW×y] | $N_{e+}$ [events/5y] | Sensitivity [$\leq \sin^2(2\theta_{13})$] |
|---|---|---|---|---|---|---|
| WCD I | 2000 | 180 | 3.8 | 7.3 | $6.8 \cdot 10^5$ | $8.0 \cdot 10^{-3}$ |
| WCD I | 2000 | 300 | 3.3 | 14.8 | $1.4 \cdot 10^6$ | $6.5 \cdot 10^{-3}$ |
| WCD + Gd | 2000 | 300 | 3.8 | 12.0 | $1.1 \cdot 10^6$ | $7.0 \cdot 10^{-3}$ |
| Scintillator | 1700 | 300 | 1.8 | 27.7 | $2.6 \cdot 10^6$ | $5.0 \cdot 10^{-3}$ |
| Scintillator∗ | 800 | 300 | 1.8 | 125.0 | $11.7 \cdot 10^6$ | $1.8 \cdot 10^{-3}$ |

At the luminosity $\mathcal{L} \geq 10^4$ t·GW·y it becomes possible to observe the distortions of spectrum caused by oscillations in the narrow range near the energy threshold (see Fig.1). In this case the distance reactor-detector will be defined practically by the threshold values of antineutrino. For the estimation of sensitivity for all detector modes except **Scintillation∗** mode we have used the average antineutrino energy. Such mode of estimation is rather obvious; however it leads to essential underestimation of experimental results at $\mathcal{L} \approx 10^4$. Thus at **Scintillation∗** mode, where the statistics allows to select with enough reliability the distortion of antineutrino spectrum in the range of 2 MeV the effective distance may be decreased from 1700m to 800m.

It follows from the table, that for all cases, excluding WCD I mode, the total number of registered positrons $N_{e+} \geq 10^6$, what corresponds to the statistical error of $\leq 0.1\%$ The total systematic error ($\sigma_{sys}$) in the scheme of experiment "one reactor - two detectors" is less than 1%. Besides the sensitivity of $\sin^2 2\theta_{13}$ measurement at the luminosity $L \geq 10^4$ [t·GW·y] weakly depends on systematic error (see Fig.1). We can conclude that the sensitivity of $\sin^2 2\theta_{13}$ measurements in LPR-SK experiment depending on detection schemes can come to $(2 \div 8) \cdot 10^{-3}$.

Conclusions

As the determinant point for the estimation of value of $\sin^2 2\theta_{13}$ parameter it's possible to consider the results of CHOOZ experiment, where the limit on $\sin^2 2\theta_{13} \leq 0.14$ has been obtained [5]. The systematic error of this experiment was 2.7%, and its value was the basic obstacle to increase the accuracy for the same experiments. For the reduction of $\sigma_{syst}$ the authors of the Kr2Det project [5] proposed a new method, so called "one reactor - two detectors" approach, which allows to remove the basic sources of errors and to reduce the total systematic error to the value of about 0.8%. It provided the relevant increase in the sensitivity of $\sin^2 2\theta_{13}$ measurements. As the following step in this direction we can consider the paper [18], where the authors performed full-scale analysis of various approaches to the reactor experiments to reach the sensitivity of $\sin^2 2\theta_{13}$ parameter < 0.01. It was shown by the authors that the influence of systematic errors on the sensitivity of the experiment is lowered due to big statistics and at luminosity ~ $10^5$ t·GW·y it is possible to reach the sensitivity to $\sin^2 2\theta_{13}$ parameter of about $10^{-3}$. For the realization of "one reactor - two detectors" experiment with such a level of luminosity it is required to construct an enormous underground detector complex just near the operating NPP with thermal power of several GW. Besides the engineering difficulties it can cause big problems connected with the radiating safety and guarding of nuclear power plant.

In the present paper it was proposed the new approach to organization of reactor experiment, which can simplify and accelerate the measurements of the $\sin^2 2\theta_{13}$ parameter with the sensitivity ≤ 0.01. The main point of proposal is the following: staying within the frameworks of "one reactor – two detectors" experimental scheme, instead of building underground detector complex near the operating NPP it is required to place a small reactor with thermal power of 300 MW at the optimal distance from the set-up with mass of about $10^4$ tons (the far detector). The experiment can be carried out at the operating Water SuperKamiokande Cherenkov detector. As SK was intended for the registration of solar neutrino and has now rather high threshold, we have analyzed some detecting methods, providing necessary sensitivity in the reactor experiments. As a result it was shown, that the use of low power reactor together with the SK detector allows to reach the sensitivity of $\sin^2 2\theta_{13} \geq (3 \div 8) \, 10^{-3}$.

For convenience we discussed experiment with the definite type of reactor KLT-40S and concrete setup – SK. However, the proposed approach assumes about free chose of reactor and neutrino detectors. For example, as "far" detector one can use KamLAND setup, at which it is possible to reach the sensitivity of $\sin^2 2\theta_{13} \leq 0.01$ without essential alteration.

The present article is dedicated to decision of only one of topical questions, and the discussion of problem how to use SPR as antineutrino source for a number of tasks of neutrino physics remained beyond its frames [30].

Acknowledgments

The authors expressed theirs sincere gratitude to L.A. Mikaelyan for very useful discussion on experimental details; to V.V.Sinev for his advices on technical features of the reactor experiments; to Mark Vagins (UCI, USA) for numerous discussions on a plan to improve the SK detector performance in the future; to Y. Kishimoto (Tohoku Univ., Japan) and Yu. Kamishkov, (Univ. of Tennessee, USA) for providing material concerning the Kamioka underground laboratory and surrounded tunnels. The authors also thank to Elena Demidova for her assistance in preparation of the paper.

This work was supported by the grant RFFR №02-02-16111.